\documentclass[11pt]{article}
\usepackage{amsfonts}
\usepackage{dsfont}
\usepackage{amsmath,amssymb}

\usepackage{bbm}
\usepackage{bm}
\usepackage{color}
\usepackage{slashed}
\usepackage{cite}
\usepackage[utf8]{inputenc}
\usepackage{graphicx}
\usepackage[normalem]{ulem}
\usepackage{bbold}
\DeclareGraphicsExtensions{.pdf,.png,.jpg}

\usepackage[colorlinks=false,citecolor=cyan,urlcolor=blue,bookmarks=true,bookmarks=true,bookmarksopen=true,bookmarksnumbered=true,bookmarksopenlevel=3]{hyperref}

\definecolor{airforceblue}{rgb}{0.36, 0.54, 0.66}
\definecolor{steelblue}{rgb}{0.27, 0.51, 0.71}
\definecolor{amber}{rgb}{1.0, 0.49, 0.0}

\renewcommand\sout{\bgroup \color[rgb]{0.55,0.00,0.99} \ULdepth=-.5ex \ULset}

\def\simg{{\ \lower-1.2pt\vbox{\hbox{\rlap{$>$}\lower6pt\vbox{\hbox{$\sim$}}}}\ }}
\def\siml{{\ \lower-1.2pt\vbox{\hbox{\rlap{$<$}\lower6pt\vbox{\hbox{$\sim$}}}}\ }}

\makeatletter \@addtoreset{equation}{section} \makeatother

\begin{document}

\flushbottom

\begin{titlepage}

\begin{centering}

\vfill

{\Large{\bf
{The hierarchy problem and fine-tuning in a decoupling approach to multi-scale effective potentials}
}} 

\vspace{0.8cm}

S.~Biondini,$^{a,}$\footnote{simone.biondini@unibas.ch} D.~Boer,$^{b,}$\footnote{d.boer@rug.nl} and R.~Peeters$^{b,}$\footnote{r.j.c.peeters@rug.nl}

\vspace{0.8cm}
 $^{a}$%
{\em Department of Physics, University of Basel,
\\
 Klingelbergstr. 82, CH-4056 Basel, Switzerland} 
\\
\vspace{0.25 cm}
$^{b}${\em Van Swinderen Institute for Particle Physics and Gravity,\\ 
University of Groningen, NL-9747 AG Groningen, The Netherlands} 
\\
\vspace{0.15 cm}

\vspace*{0.8cm}

\end{centering}

\vspace*{0.3cm}
 
\noindent

\begin{abstract}
In many realizations of beyond the Standard Model theories, new massive particles are introduced, leading to a multi-scale system with widely separated energy scales. In this setting the Coleman-Weinberg effective potential, which describes the vacuum of the theory at the quantum level, has to be supplemented with a prescription to handle the hierarchy in mass scales.  In any quantum field theory involving scalar fields and multiple, highly differing mass scales, it is in general not possible to choose a single renormalization scale that will remove all the large logarithms in the effective potential. In this paper, we focus on the so-called decoupling method, which freezes the effects of heavy particles on the renormalization group running of the light degrees of freedom at low energies.  We study this for a simple two-scalar theory and
find that, while the decoupling method leads to an acceptable and convergent effective potential, the method does not solve the fine-tuning problem that is inherent to the hierarchy problem of multi-scale theories. We also consider an alternative implementation of the decoupling approach, which gives different results for the shape of the potential, but still leads to similar conclusions on the amount of fine-tuning in the model.  We suggest a way to avoid running into this fine-tuning problem by adopting a prescription on how to fix parameters in such decoupling approaches.

\end{abstract} 
\vfill

\newpage
\end{titlepage}
\section{Introduction}
The Standard Model (SM) of particle physics is our present-day reference for a detailed understanding and description of the fundamental building blocks of matter and the interactions among them. Despite its success, some new physics is expected because of various compelling observations and measurements that still require clarification, such as the nature and generation of neutrino masses, the hierarchies among fermion masses, the nature of dark matter, the baryon asymmetry in the universe and the origin of electroweak symmetry breaking. The current view of the SM, which has developed over the past few decades, is that of an effective low-energy theory of some unknown ultraviolet (UV) completion. 

If the theory that extends the SM is meant to solve one of the aforementioned open questions, it is likely that new degrees of freedom couple with the SM particles. The Higgs sector is especially interesting in this regard, as the Standard Model Higgs boson is often taken as a portal between the visible and a dark sector \cite{McDonald:1993ex,Burgess:2000yq,Schabinger:2005ei,Feng:2014vea,Wang:2015cda,Steele:2013fka}, it enters the interactions with right-handed neutrinos in the leptogenesis framework to account for the baryon asymmetry \cite{Fukugita:1986hr}, and the origin of the electroweak symmetry breaking could be explained if the Higgs couples to additional scalar particles \cite{Espinosa:2007qk,Englert:2013gz,Farzinnia:2013pga,Khoze:2013uia}.   
However, the existence of interactions between the Higgs boson and any heavier state of new physics comes at the price of generating quantum corrections to the Higgs mass that are quadratic in the mass of the heavy particle \cite{Gildener:1976ai,Susskind:1978ms,tHooft:1979rat,Weinberg:1978ym,Veltman:1980mj}.
These corrections cannot be avoided on the basis of symmetry arguments and make the Higgs mass extremely sensitive to high-scale physics, which is viewed as unnatural \cite{tHooft:1979rat,Giudice:2013yca}. 
This constitutes the so-called \textit{hierarchy problem}: how to explain why the electroweak scale is so small if there is any beyond the Standard Model physics at high scales.

The hierarchy problem can also be viewed as a fine-tuning problem, as the parameters of new high-scale physics have to be chosen very carefully in order to result in the observed low-energy parameters. Besides in the vacuum energy density, this effect is most pronounced in the mass parameter, which receives large corrections in the renormalization group (RG) running from high to low energies when there is a large hierarchy between the different scales. There are some well-known theories that address the hierarchy problem by introducing new physics at the TeV scale, like supersymmetric extensions \cite{Wess:1973kz,Iliopoulos:1974zv,Kaul:1981wp,Martin:1997ns}, composite Higgs \cite{Kaplan:1983sm, Dugan:1984hq,Bellazzini:2014yua} and extra dimensions \cite{ArkaniHamed:1998nn, ArkaniHamed:1998rs, Randall:1999ee}. However the minimal versions of these models are nowadays rather fine-tuned in order to stay compatible with experiments \cite{Giudice:2013yca,Feng:2013pwa,Wells:2013tta,Barnard:2015ryq}.

In this paper we do not consider specific theories that aim to solve the hierarchy problem, but will take a closer look at various aspects of the hierarchy problem from the perspective of the effective potential, also referred to as the Coleman-Weinberg (CW) potential \cite{Coleman:1973jx}.\footnote{See ref.\ \cite{Quiros:1999jp} for an extensive review of the effective potential both at zero and finite temperature.} The CW potential describes the vacuum of the theory at the quantum level, i.e.\ the true minimum of the theory. At the level of the CW potential the hierarchy problem shows up through large loop corrections. It is well-known that in any quantum field theory involving scalar fields and multiple, highly differing mass scales, it is in general not possible to choose a single renormalization scale that will remove all the large logarithms in the effective potential and as a consequence, in its derived quantities, such as the vacuum expectation value (VEV) and the mass of the low-energy scalar field. To be specific, the loop expansion of the effective potential contains logarithms of the ratio $m_i^2/\mu^2$, with $m_i$ being the mass eigenvalues of the model, and $\mu$ the renormalization scale. In a model with a hierarchy in mass scales, it is impossible to choose a value for $\mu$ such that all logarithms containing the ratio $m_i^2/\mu^2$ are small at the same time. The remaining large logarithms will appear with higher powers at higher orders, invalidating the loop expansion. The non-convergence of the perturbative expansion of the effective potential in the presence of a multi-scale system will hamper its predictiveness.

To address this problem various approaches to the CW potential of multi-scale systems have been put forward: (i) multi-scale renormalization methods \cite{Einhorn:1983fc,Ford:1994dt,Ford:1996yc,Steele:2014dsa}; (ii) a decoupling method in mass-independent renormalization schemes \cite{Bando:1992wy,Bando:1992np,Casas:1998cf}; (iii) a single-scale renormalization-group improvement for multi-scale effective potentials~\cite{Chataignier:2018aud}; (iv) effective field theory (EFT) techniques \cite{Masina:2015ixa,Manohar:2020nzp}. As emphasized in \cite{Manohar:2020nzp} one should not confuse an EFT approach to calculating the CW potential with calculating the CW potential of a low energy EFT. In \cite{Manohar:2020nzp} the former is done, whereas in \cite{Masina:2015ixa} the CW potential of the low energy EFT is matched onto the CW potential of the full theory at the scale of the high mass modes. In \cite{Manohar:2020nzp} it was shown that the EFT approach reproduces the CW potential and the accompanying matching conditions in the decoupling method for the Higgs-Yukawa model \cite{Casas:1998cf} at the two-loop level. In ref.\ \cite{Casas:1998cf} it was furthermore shown within the Higgs-Yukawa model that the decoupling method is equivalent to the multi-scale method used in \cite{Ford:1994dt,Ford:1996yc}. In contrast, the single-scale improvement is not equivalent to the other methods, as it does not always resum the leading logarithms \cite{Chataignier:2018aud}. 

Each of these approaches proposes a way to handling the heavy modes and the large quantum corrections in order to arrive at a predictive/convergent effective potential. For each of these approaches one may wonder whether the hierarchy problem still manifests itself or that has been solved at the level of the effective potential at least. In the approach of \cite{Masina:2015ixa} the hierarchy problem in its original form appears in the matching condition for the low-energy scalar mass, where one has to require a very fine cancellation between two large quantities when imposing the lightness of the mass of a scalar $\phi$ at a high energy scale $\mu_S$ (in our notation): 
\begin{equation}
    m^2_{\phi,\hbox{\tiny LE}}(\mu_S)=m_{\phi}^2(\mu_S) - \frac{\alpha^2 }{8 \pi} \mu_S^2,
    \label{mass_EFT0}
\end{equation}
where $m_{\phi,\hbox{\tiny LE}}$ is the $\phi$-scalar mass in the low-energy theory, whereas $m_\phi$ is the mass in the high-energy theory.
Slightly changing the value of the high-energy parameter $m_\phi$ at $\mu> \mu_S$ where the heavy modes contribute, and running it down to $\mu=\mu_S$, would have a significant impact on Eq.~\eqref{mass_EFT0} and it would spoil the lightness of $m_{\phi,\hbox{\tiny LE}}$. Note that in this expression the significant impact is due to the quadratic correction, not due to the large evolution step, i.e.\ $\mu/\mu_S$ does not need to be very large. This is different from the logarithmic case which is usually deemed acceptable (e.g.\ small changes of $\alpha_s$ at $M_Z$ do not lead to big changes at the GeV scale due to the logarithmic running). One concludes that the hierarchy problem is present as a fine tuning problem in the matching condition. In the EFT approach of \cite{Manohar:2020nzp} this fine-tuning problem shows up in the matching condition for the effective potential, but is avoided by selecting a convenient scale. It is chosen such that the high energy and low energy fields do not mix at this scale. The hierarchy problem for the energy density cannot be removed at the same time though. Therefore, some remnant of the hierarchy problem remains, but predictions for the quantities derived from the CW potential can be made free of any fine-tuning upon choosing the right matching scale.

In this paper we investigate this same issue for the decoupling method of \cite{Casas:1998cf}, which to some level agrees with the approach of \cite{Manohar:2020nzp}. For concreteness we consider a simple model with two scalar fields, where we take the mass parameters with a large scale separation. Moreover, the light scalar field undergoes spontaneous symmetry breaking (SSB) at tree-level, whereas the additional heavy scalar field, that mimics the new physics, does not (see refs.~\cite{Casas:2000mn,Masina:2015ixa} for a similar model and for a model where both scalars undergo SSB see \cite{Manohar:2020nzp}). The two-scalar model comprises the main ingredients to inspect fully the hierarchy problem and it may resemble the Standard Model Higgs sector extended with an unknown high-energy completion. Our goal is to address two main aspects of the hierarchy problem when implementing the decoupling approach: (i) checking whether the SSB still occurs at the quantum level in the presence of high-energy degrees of freedom, and to which extent the minimum of the potential  and the mass of the light scalar are affected; (ii) inspecting the fine-tuning of the model parameters. 

The paper is organized as follows. In Sec.~\ref{sec:potential_no_dec} we introduce the two-scalar model and highlight the different aspects of the hierarchy problem in the context of the effective potential. Then in Sec.~\ref{sec:decoupling}, after the decoupling method is introduced, we address the numerical evaluation of the VEV of the low-energy scalar field, its stability under radiative corrections and its mass in Sec.~\ref{sec:minimum_dec}, while in Sec.~\ref{sec:finetuning} the connection between fine-tuning and the effective potential is studied. In Sec.~\ref{sec:comparison} we compare to a different implementation of the decoupling method put forward in the literature and consider a prescription to fix parameters in order to avoid the fine-tuning problem. Finally,  we summarize our conclusions in Sec.~\ref{sec:concl}.

\section{A large hierarchy in the effective potential}
\label{sec:potential_no_dec}
In order to set a reference for the decoupling approach, we discuss the different problems that appear when we naively apply the effective potential formalism to a theory with a hierarchy in the mass parameters (referred to as the non-decoupling approach). To this end, we work with a simple model, which is a theory with two real scalar particles. The model Lagrangian can be written as follows:
\begin{eqnarray}
\mathcal{L} = \frac{1}{2} \partial_\mu \phi \, \partial^\mu \phi + \frac{1}{2} \partial_\mu S \, \partial^\mu S - V(\phi,S) \, , 
\label{Lag_model}
\end{eqnarray}
where the potential with bare parameters reads
\begin{eqnarray}
V(\phi,S) = \frac{1}{2}\mu_\phi^2\phi^2 + \frac{1}{2} \mu_S^2 \, S^2 + \frac{\lambda_\phi}{4}\phi^4 + \frac{\lambda_S}{4}S^4 + \alpha \, \phi^2 S^2 \, .
\label{Bare_potential}
\end{eqnarray}
On the one hand, the parameters of the field $\phi$ are chosen to achieve spontaneous symmetry breaking at tree level, so we take $\mu_\phi^2 < 0$ with the corresponding VEV $v_\phi= 
\pm\sqrt{-\mu_\phi^2/\lambda_\phi}$. On the other hand, the mass parameter for the $S$ field already corresponds to a physical mass and we take it to 
satisfy $\mu_S^2 \gg |\mu_\phi^2|$. As mentioned, the scalar field $S$ plays the role of a high-energy degree of freedom and it belongs to the UV energy domain in our model (the highest scale we shall consider when running the parameters will be $\mu_S$). We remark that the heavy field does not develop a VEV. This will still hold under radiative corrections to the heavy scalar mass, since the corrections from both $\mu_S$ and $\mu_\phi$ are not large enough to flip the sign of $\mu_S^2$.
To ensure that the potential Eq.~\eqref{Bare_potential} is bounded from below, there are some constraints on the quartic couplings (see e.g.~ref.~\cite{Kannike:2012pe}):
\begin{equation}
    \lambda_\phi > 0 \, , \quad \lambda_S > 0 \, , \quad \alpha > -\frac{1}{2}\sqrt{\lambda_\phi \lambda_S} \, .
\end{equation}

Next we turn to the effective potential for the model. The RG-improved effective potential in any mass-independent renormalization scheme can be organized in a loop expansion $ V_{\hbox{\scriptsize eff}}=V^{(0)}+V^{(1)}+\dots$, where $V^{(0)}$ is the RG-improved tree-level potential and $V^{(1)}$ is the one-loop correction after renormalization which reads (using the $\overline{\text{MS}}$ scheme):
\begin{equation}
    V^{(1)} = \frac{1}{4 (4\pi)^2}\left\{ m_\phi^4\left[\log\frac{m_\phi^2}{\mu^2}-\frac{3}{2}\right] + m_S^4\left[\log\frac{m_S^2}{\mu^2}-\frac{3}{2}\right]
    \right\}.
    \label{eq:Veff_ren}
\end{equation}
 The masses in this expression are the tree-level masses, as a function of the classical field value $\phi_c$. They are found to be
\begin{equation}
    m^2_\phi= \mu_\phi^2 + 3\lambda_\phi \phi_c^2\, , \quad m^2_S = \mu_S^2  + 2\alpha \, \phi_c^2 \, .
    \label{eq:mass_eigen}
\end{equation}
\begin{figure}[t!]
    \centering
    \includegraphics[scale=0.8]{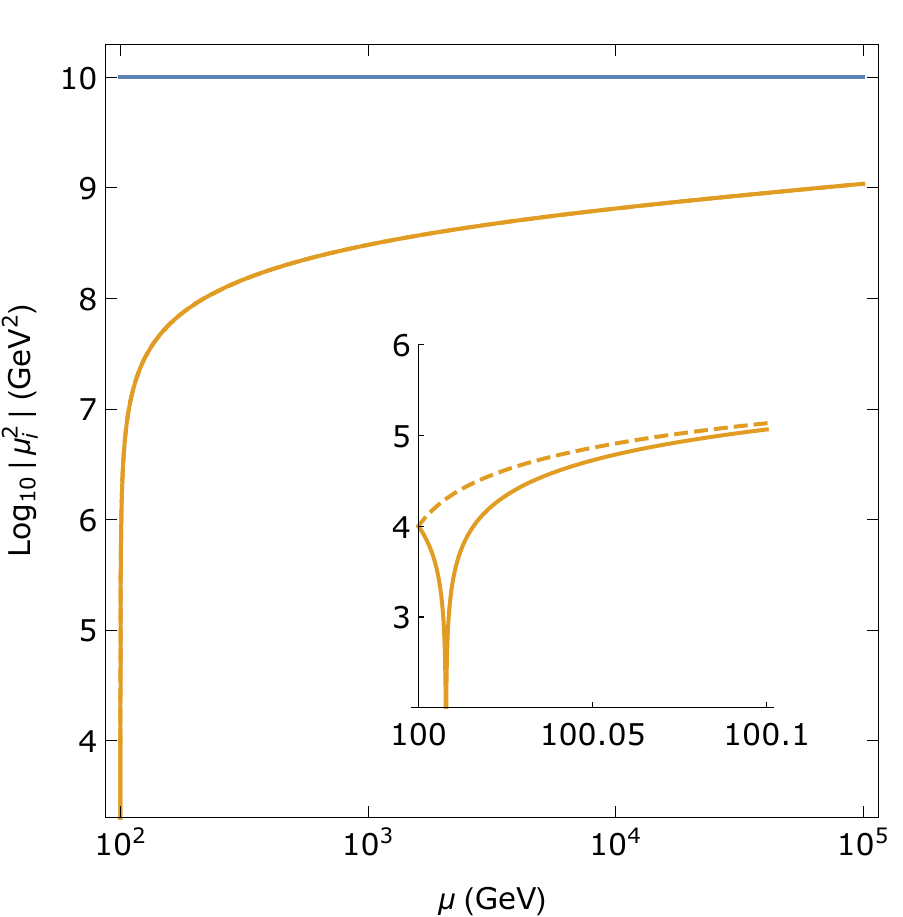}
    \caption{The RG running of the mass parameter $\mu_\phi^2$ (orange) at 1-loop level using the benchmark point described in the text, with $\alpha > 0$ (solid line) and $\alpha < 0$ (dashed line). The value of $\mu_S^2$ is constant (solid blue), as explained in the text.}
    \label{fig:RG_NonDecoupling}
\end{figure}
According to our assumptions, the heavy scalar does not develop a nonzero VEV by construction, so that we can set $S_c=0$. Therefore, the masses in Eq.~(\ref{eq:mass_eigen}) are independent of $S_c$. When extracting the counterterms of the effective potential, one can see that $\mu_S^2$, $\lambda_S$ and $\alpha$ do not run, in contrast to $\lambda_\phi$ and $\mu_\phi^2$. The resulting $\beta$-functions for $\lambda_\phi$ and $\mu_\phi^2$ are given by:\footnote{We checked that, when letting $S_c$ have a varying field value, we recover the full set of RGEs, see e.g.~ref.~\cite{Ookane:2019iwq}.}
\begin{align}
    \beta_{\lambda_\phi} = \mu \frac{\partial \lambda_\phi}{\partial \mu} &= \frac{1}{8\pi^2}(9\lambda_\phi^2 + 4\alpha^2) \, ,
    \label{RGEs_no_dec_1}
    \\
    \beta_{\mu_\phi^2} = \mu \frac{\partial \mu_\phi^2}{\partial \mu} &= \frac{1}{8\pi^2}(3\lambda_\phi^{\phantom{2}} \mu_\phi^2 + 2\alpha \mu_S^2) \, .
    \label{RGEs_no_dec_2}
\end{align}
In order to illustrate the behavior of this model, we will use a benchmark point for the parameter space throughout the paper. The parameter values are:
\begin{align}
\begin{split}
&\mu_\phi^2(\mu_\text{min}) = -(10^2\text{ GeV})^2 \, , \quad \mu_S^2(\mu_\text{min}) = (10^5\text{ GeV})^2 \, , 
\\
&\lambda_\phi(\mu_\text{min}) = 0.8 \, , \quad  \lambda_S(\mu_\text{min}) = 1.3 \, , \quad  \alpha(\mu_\text{min}) = \pm 0.5 \, .
\end{split}
\label{banch}
\end{align}
where $\mu_\text{min} = 100 \text{ GeV}$.\footnote{Note that this does not mean that the parameters values are fixed at the scale 
$\mu_\text{min}$, but just that they take these values at that scale. } Furthermore, we take $\mu_S$ as the maximum value for the renormalization scale, so $\mu_\text{max} = 10^5$ GeV. As we will see, changing the sign of $\alpha$ has a large effect on some of the results. Hence, we will use both a positive and negative value for $\alpha$ in our benchmark point. We adopt this set of parameters for all numerical results presented in our paper, unless explicitly stated otherwise. The running of the mass parameter $\mu_\phi^2$ (and the comparison with the constant $\mu_S^2$) in this model is shown in Fig.~\ref{fig:RG_NonDecoupling}. One observes that $\mu_\phi^2$ receives large corrections from the high-energy modes already for small deviations from the initial renormalization scale. This is an expression of the hierarchy problem. When $\alpha > 0$, the $\mu_S^2$ contributions to the running of $\mu_\phi^2$ are positive, and $\mu_\phi^2$ flips sign already at $\mu = 100.008$ GeV (solid orange line). At high energy a large positive value of $\mu_\phi^2 \approx 10^9$ GeV is reached. For $\alpha < 0$, there are only negative contributions to the running of $\mu_\phi^2$. So the value at high energy will have the opposite sign as in the case $\alpha > 0$, but the absolute value is essentially the same, since the small initial value of $\mu_\phi^2$ at $\mu_\text{min}$ has practically no effect on the high-energy value. 

\begin{figure}[t!]
    \centering
    \includegraphics[scale=0.7]{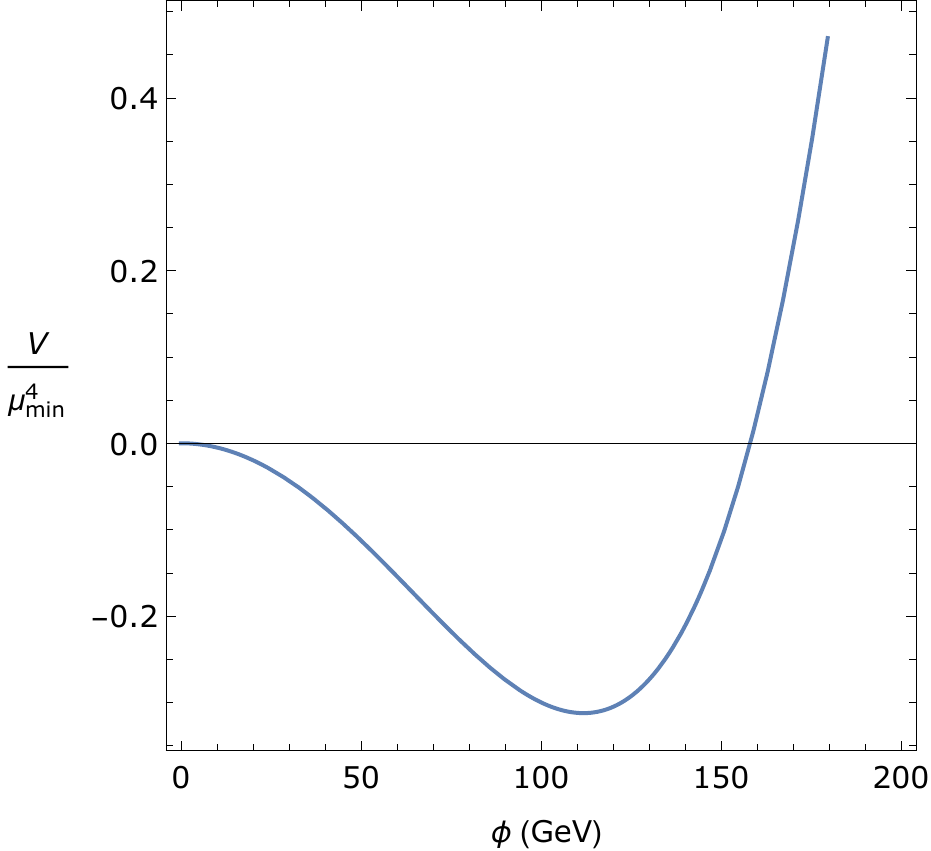}
    \hspace{0.8 cm}
    \includegraphics[scale=0.7]{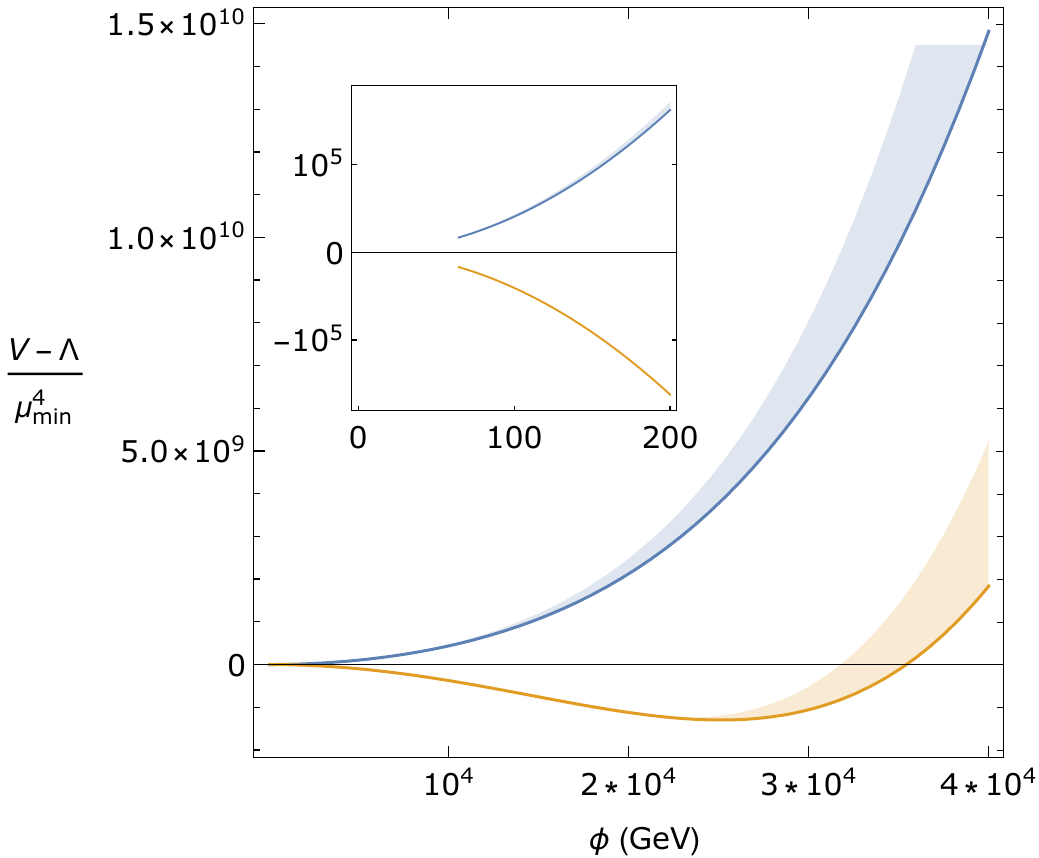} 
    \caption{The shape of the scalar potential in the non-decoupling case. The left figure shows the tree-level result, evaluated at $\mu = \mu_\text{min}$. The right plot shows the results including the 1-loop contributions for $\alpha > 0$ (blue solid line) and $\alpha < 0$ (orange solid line). In the right figure,  the vacuum energy or cosmological constant $\Lambda$ is subtracted as explained in the text to normalize the potential to zero at $\phi = 0$. The solid lines in the right plot show the result for $\mu = \mu_\text{min}$, and a band is shown for the range $\mu_\text{min} \leq \mu \leq \mu_\text{max}$.}
    \label{fig:ShapePotential_NonDecoupling}
\end{figure}
\subsection{Properties of the light scalar from the effective potential}
\label{sec:properties}
In this section we address the shape of the potential and the corresponding mass eigenvalue of the scalar field $\phi$ when we include the 1-loop corrections. Let us begin with the study of the shape of the potential for different values of the renormalization scale, and for both $\alpha > 0$ and $\alpha < 0$, that we show in Fig.~\ref{fig:ShapePotential_NonDecoupling}.
The left plot shows the tree-level result evaluated at the scale $\mu_\text{min}$. This potential exhibits the well-known Mexican-hat behavior. In the right plot, the 1-loop contributions are added and the outcome is quite different from the tree-level result. First of all, as can be seen from eqs.~(\ref{eq:Veff_ren}) and (\ref{eq:mass_eigen}), there is a large constant in the potential ($\propto \mu^4_S$), that contributes to the vacuum energy, or cosmological constant $\Lambda$, that we have set to zero in \eqref{Lag_model}. In order to normalize the potential to zero at $\phi = 0$, and better compare the shape of the potential with the tree-level case, we subtract from the one-loop potential the cosmological constant 
\begin{equation}
 \Lambda = \frac{\mu_S^4}{64\pi^2}\left( \log\frac{\mu_S^2}{\mu^2} -\frac{3}{2} \right) + \frac{\mu_\phi^4}{64\pi^2}\left( \log\frac{\mu_\phi^2}{\mu^2} -\frac{3}{2} \right).
\label{sub_c_term}
\end{equation}
We note that the $\mu_\phi^4$ term in Eq.~\eqref{sub_c_term} is generally subdominant. Looking at the shape of the potential, there is a clear difference between $\alpha > 0$ and $\alpha < 0$. When $\alpha > 0$, there is no non-trivial minimum anymore, hence no spontaneous symmetry breaking can occur (solid blue line). This happens because the contributions proportional to $m_S^4$ dominate the potential, and they are all positive (recall the expression for $m_S^2$ in Eq.~(\ref{eq:mass_eigen})). In the case $\alpha < 0$, the $m_S^4$ term in the potential gives both positive ($\alpha^2\phi^4$) and negative ($\alpha \mu_S^2 \phi^2$) contributions. The competition between these terms results again in a Mexican-hat shape, as can be seen in the right panel of Fig.~\ref{fig:ShapePotential_NonDecoupling} (solid orange line). However, the location of the VEV is very different from the tree-level case, with a value for the 1-loop VEV of $v_\phi^{(1)} \approx 2.5 \cdot 10^4$ GeV much larger than the tree-level VEV $v_\phi^{(0)} \approx 112$ GeV. In the figure, a band is shown for the range $\mu_\text{min} \leq \mu \leq \mu_\text{max}$ which demonstrates that the results are largely independent of the chosen scale. We also note that for small values of $\phi = \phi_c$, both lines stop, which happens because of the appearance of imaginary values of the potential. This is a well-known feature that does not imply a breakdown of the effective potential approach, but rather signals a decaying state. For details we refer to \cite{Weinberg:1987vp}. 

Next, we want to see the effect of the 1-loop contributions on the mass of the light boson, as this is important in understanding the hierarchy problem in this model. In general, the mass of $\phi$ is given by the second derivative of the potential with respect to $\phi$. The full expression contains the tree-level contribution and 1-loop contributions from the $\phi$ and $S$ sectors. We single out the dominant term, which is the one that depends on the heavy scale $\mu_S^2$:
\begin{equation}
    m_{\phi}^{2,(1)} (\mu) \sim \frac{\alpha}{16\pi^2}(\mu_S^2 + 2\alpha\phi_c^2)\left[\log \frac{\mu_S^2 + 2\alpha\phi_c^2}{\mu^2}-1\right].
    \label{eq:mass_1loop_full}
\end{equation}
For $\mu \not\approx e^{-1/2}m_S$, this term is large and pushes the mass of $\phi$ to large values. Notably, there is a large shift in $m_\phi^2$ already for $\mu = \mu_\text{min}$. This is how the hierarchy problem of the Higgs mass manifests itself in this model.

As mentioned in the introduction, the perturbative expansion of the effective potential involves different large logarithms when a large hierarchy in scales is present. As can be seen from the expression for the 1-loop potential in Eq.~(\ref{eq:Veff_ren}), ratios of scales appear in the two logarithms: $m^2_\phi/\mu^2$ and $m^2_S/\mu^2$ with $m^2_\phi \ll m^2_S$. Because of this hierarchy it is not possible to find a value of $\mu$ such that both logarithms in Eq.~(\ref{eq:Veff_ren}) are small at the same time. The loop expansion is invalidated and the CW potential of the Lagrangian \eqref{Lag_model} is not predictive: we see that the effect of including the 1-loop corrections is large for the shape of the potential, the VEV of the low-energy field $\phi$ and the mass of $\phi$. Even at the scale $\mu = \mu_\text{min}$, there are large corrections to the location of the VEV and the mass of the light scalar. The decoupling method has been proposed as a possible solution to this problem and in the next section we implement this approach. 
\section{The decoupling method}
\label{sec:decoupling}
The main idea behind the decoupling method \cite{Bando:1992np,Bando:1992wy,Casas:1998cf} is to decouple any particle in the theory as soon as the energy is too low to excite its modes \cite{Appelquist:1974tg}. In practice, one introduces step functions in the effective potential, such that the contribution of particles with a mass larger than the decoupling scale $\mu_{\hbox{\scriptsize dec}}$ is switched off.  Using this prescription, the full effective potential can be safely evaluated down to the energy scale of the lightest particle.
For the model at hand, the one-loop contribution can then be written as follows
\begin{equation}
    V^{(1)}_{\hbox{\scriptsize dec}} = \frac{1}{4 (4\pi)^2}\left\{ \theta(\tilde{\mu}-m_\phi)  m_\phi^4  \log\frac{m_\phi^2}{\tilde{\mu}^2} \,  + \theta(\tilde{\mu}-m_S) m_S^4\log\frac{m_S^2}{\tilde{\mu}^2} \,  
    \right\} \, ,
    \label{eq:Veff_ren_dec} 
\end{equation}
where we write $\tilde{\mu}^2 = \mu^2 e^{3/2}$, in order to absorb the usual numerical factor, which is $3/2$ for scalars. We can determine the $\beta$-functions by taking the derivative with respect to the renormalization scale $\mu$ of $ V^{(1)}_{\hbox{\scriptsize dec}}$.\footnote{We note that the step functions do not contribute to this derivative, since they give a factor $m_i^4\delta(\tilde{\mu}-m_i)\log\frac{m_i^2}{\tilde{\mu}^2}$. This factor is always zero, because for the single value where the $\delta$-function is non-zero, the logarithm vanishes. Also note that one could replace the step functions by smooth versions, but this will not affect our conclusions.} In order to avoid some clutter, we define $\theta_i \equiv \theta(\tilde{\mu}-m_i)$. 
The decoupling is inherited by the two parameters $\lambda_\phi$ and $\mu_\phi^2$ so that the $\beta$-functions now read:
\begin{align}
    \beta_{\lambda_\phi} = \mu \frac{\partial \lambda_\phi}{\partial \mu} &= \frac{1}{8\pi^2}(9\lambda_\phi^2\theta_\phi^{\phantom{2}} + 4\alpha^2\theta_S^{\phantom{2}}) \, ,
    \label{run_dec_lambda}
    \\
    \beta_{\mu_\phi^2} = \mu \frac{\partial \mu_\phi^2}{\partial \mu} &= \frac{1}{8\pi^2}(3\lambda_\phi^{\phantom{2}} \mu_\phi^2\theta_\phi^{\phantom{2}} + 2\alpha \mu_S^2\theta_S^{\phantom{2}}) \, .
    \label{run_dec_muphi}
\end{align}
\begin{figure}[t!]
    \centering
    \includegraphics[scale=0.73]{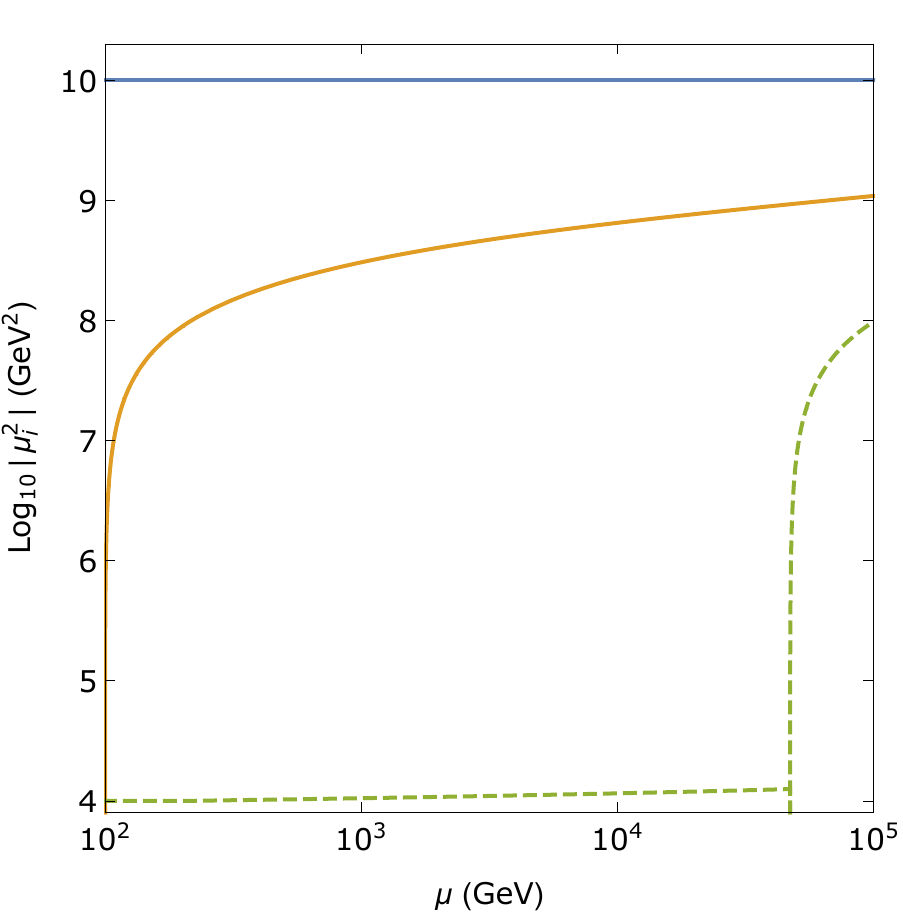}
    \hspace{0.3 cm}
     \includegraphics[scale=0.77]{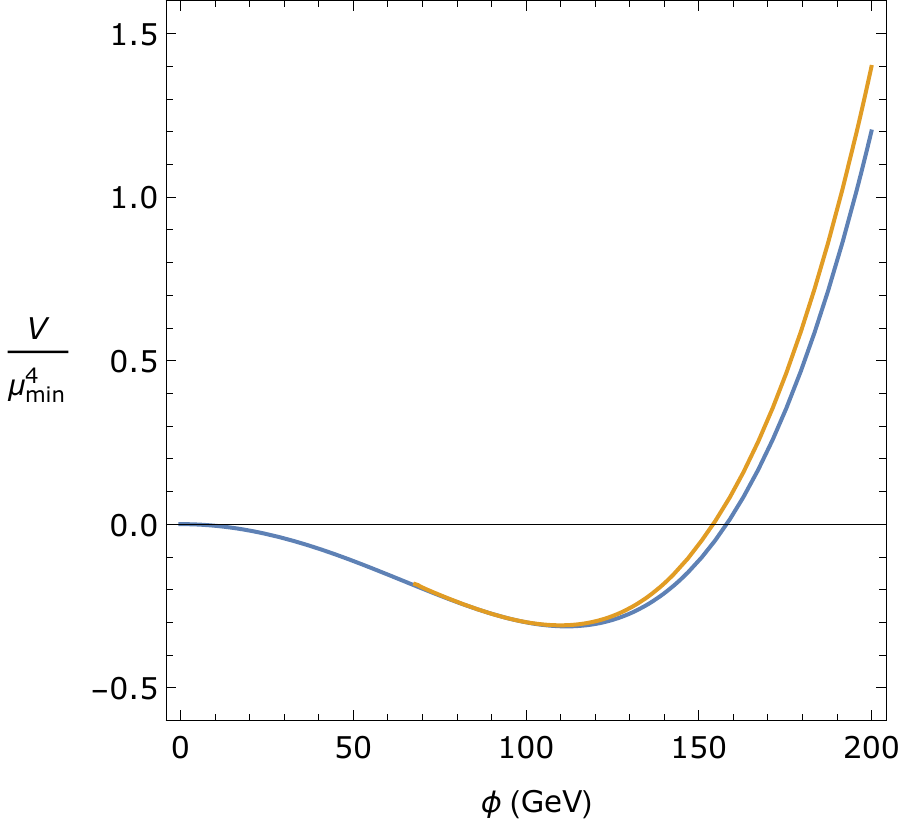}
    \caption{(Left) The running of the mass parameters, with $\mu_S^2$ in blue, and $\mu_\phi^2$ in solid orange for the case without decoupling, and in dashed green for the case with decoupling, using $\alpha > 0$. (Right) A comparison between the shape of the tree-level potential evaluated at $\mu = \mu_\text{min}$ (blue line) and the tree-level + 1-loop potential using the decoupling method (orange line). The potential is plotted in units of $V/\mu_\text{min}^4$.}
    \label{fig:RGrunningCombined}
\end{figure}

In our setting, when we run down from high energy, we obtain first a decoupling of the heavy $S$ field.  For $\tilde{\mu} < m_S$, i.e.\ $\mu \approx \mu_\text{max}/2$, the effect of the heavy degree of freedom on the running parameters is absent and the change with the renormalization scale is only due to the lighter scalar $\phi$. Going further down $\tilde{\mu} < m_\phi$, the couplings are frozen and do not run anymore. So the decoupling method provides a prescription for the renormalization scale $\mu$ \cite{Casas:1998cf}, which is setting it equal to or smaller than the lowest decoupling scale $\mu^* \equiv m_{\phi} \, e^{-3/4}$ (which in our setting we take anyway larger than $\mu_{\textrm{min}}$). In this way, the loop contributions are absent, so the problematic logarithms disappear from the potential. Only the RG-improved tree-level contribution $V^{(0)}(\mu < \mu^*)$ remains. This prescription ensures that higher-loop contributions also vanish, restoring the predictivity of the perturbative expansion.\footnote{In order to find an appropriate renormalization scale, the authors of ref.~\cite{Casas:1998cf} suggested two conditions: $V^{(1)}_{\hbox{\scriptsize dec}}(\mu)=0 $ (best perturbative convergence) and  $d(V^{0}+V^{(1)}_\textrm{dec})/d\mu$=0 (least $\mu$-dependence). When choosing the renormalization scale  smaller than the smallest mass eigenvalues, the two prescriptions are fully equivalent.}

In Fig.~\ref{fig:RGrunningCombined} (left) one finds the mass parameter $\mu_\phi^2$ evolved with and without decoupling, in order to show the strongly different behavior. The same benchmark point is used as in the case without decoupling, with $\alpha > 0$. This figure shows that when using the decoupling method, $\mu_\phi^2$ remains small for most of the considered energy range since the high-energy modes only start contributing at $\mu \approx \mu_\text{max}/2$. Moreover, although the high-energy modes do contribute to the RG running for $\tilde{\mu} > m_S$, the final value of $\mu_\phi^2$ at $\mu_\text{max}$ is one order of magnitude lower than in the case without decoupling. The results are essentially the same for $\alpha < 0$, even though it changes the value of $m_S$ (cf.\ Eq.~\eqref{eq:mass_eigen}) and therefore the value of the decoupling scale. This effect will be small, because we only consider small values for $\phi_c$, so it will not change the fact that the high-scale modes contribute above $\mu \approx \mu_\text{max}/2$.
\subsection{Properties of the potential in the decoupling approach}
\label{sec:minimum_dec}
Let us now use the prescription for the renormalization scale in the decoupling method to evaluate the effective potential. A comparison between the tree-level result and the one-loop result in the decoupling method is shown in Fig.~\ref{fig:RGrunningCombined} (right). It is worth stressing that the shape of the 1-loop effective potential (solid-orange line in Fig.~\ref{fig:RGrunningCombined}) is independent of the sign of $\alpha$ (whereas this affects severely the 1-loop effective potential in the naive implementation, see Fig.~\ref{fig:ShapePotential_NonDecoupling}). This is due to the absence of $\alpha$-dependent terms at scales $\tilde{\mu} < m_S$. In contrast to the case without decoupling, we now see that the 1-loop effective potential only introduces small corrections, signalling a good perturbative expansion at least in the low-energy domain and at this loop order. Moreover, one can see that a non-trivial minimum is realized and spontaneous symmetry breaking can occur, also for $\alpha > 0$. The truncation of the orange solid line is due to the squared mass eigenvalue $m_\phi^2$ turning negative for values $\phi_c \leq 68$ GeV for this choice of the parameters. This makes it impossible to define a meaningful decoupling scale, and leads to the aforementioned decaying state~\cite{Weinberg:1987vp}, but we will not address this further here. 

We have checked numerically that the good agreement between the tree-level potential and the decoupling potential is not due to the specific choice of parameters in the benchmark point. For different low-energy values of $\lambda_\phi$ and $\mu_\phi^2$, the shift in the value of the VEV $v^{(1)}$ is at the few-percent level.

Let us now look at what happens to the mass of the $\phi$ scalar when the decoupling method is used. Since the 1-loop contributions are absent from the potential in this case ($V_\textrm{dec}^{(1)}=0$), the term that previously pushed $m_\phi$ to large values no longer contributes. The mass of the $\phi$ boson is now given by:
\begin{equation}
    m_{\phi}^2(\mu < \mu^*) = \mu_\phi^2 + 3\lambda_\phi\phi_c^2,
    \label{mass_phi_dec}
\end{equation}
where all parameters are understood to be evaluated at the same scale $\mu <\mu^*$. In contrast to Eq.~(\ref{eq:mass_1loop_full}), there are no $\mu_S^2$ contributions in Eq.~(\ref{mass_phi_dec}), and the decoupling method ensures the lightness of the $\phi$ boson.

\section{Fine-tuning of the model}
\label{sec:finetuning}
The decoupling approach comes as a solution for the issues related to the perturbative expansion of the CW potential, and shields the mass parameter from large corrections by choosing the renormalization scale below the decoupling scale. The next step is to investigate whether there is still fine-tuning present when this method is used. As mentioned in the introduction, the matching conditions can be a source of fine-tuning  in the EFT approach. We want to see whether and how the fine-tuning appears in the non-decoupling approach.  
In practice we shall make contact between the effective potential, with and without the decoupling of the field $S$ implemented, and the amount of fine-tuning that is necessary at high energy ($\mu_\text{max}$) in order to obtain a given \emph{small} value for $\mu_\phi^2$ at low energy ($\mu_\text{min}$). 

\begin{figure}[t!]
    \centering
    \includegraphics[scale=0.62]{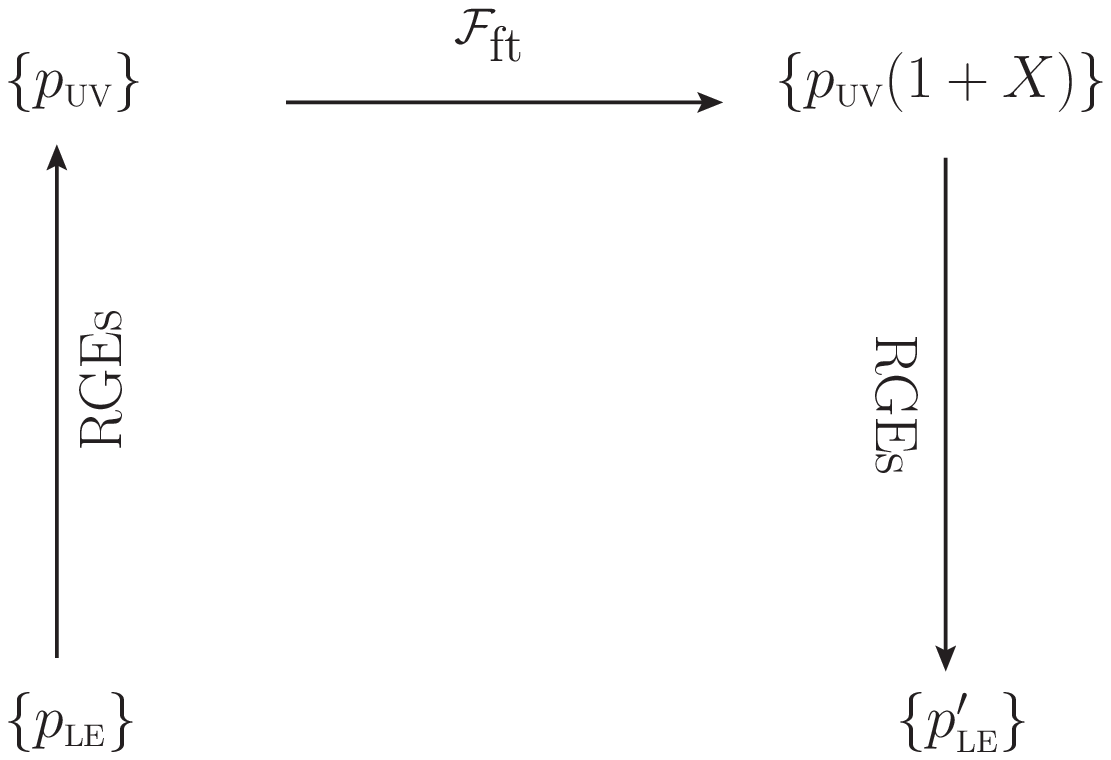}
    \hspace{0.6 cm}
     \includegraphics[scale=0.8]{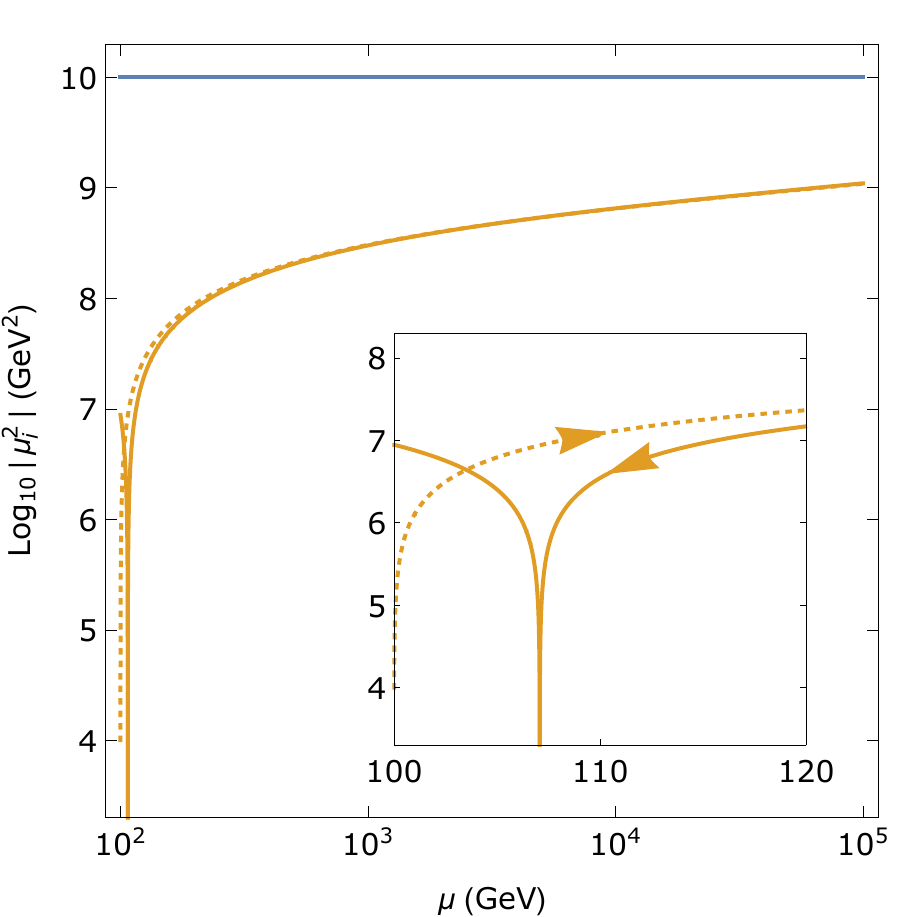}
    \caption{(Left) Diagrammatic representation of the prescription for determining the amount of fine-tuning ($p_{\textrm{LE}}$ and $p_{\textrm{UV}}$ stand for the parameters at low-energy and at the ultraviolet scale respectively). There is running of the parameters from low to high energy, a shift of the parameters through $\mathcal{F}_\text{ft}$ and running back to low energy. (Right) The result of running the mass parameters up from low energy (dotted line) and down from high energy after changing the high-energy parameters with $X$ = 0.01 (solid line), with $\alpha < 0$.}
    \label{fig:FTrunning_NonDecoupling}
\end{figure}

To this end, we start with the model parameters specified at the initial scale $\mu_\text{min}$.  Then, we run the parameters up to the high-energy scale $\mu_\text{max}$ using the RG equations (\ref{RGEs_no_dec_1}) and (\ref{RGEs_no_dec_2}). Here, we implement a variation by multiplying the parameters with $\mathcal{F}_{\hbox{\scriptsize ft}}=1+X$, with $X \in [0,1]$, and then run the parameters back down to $\mu_\text{min}$. A diagrammatic representation of the prescription is given in Fig.~\ref{fig:FTrunning_NonDecoupling} (left). 
We use this same procedure for both the decoupling and non-decoupling approach. As an example of the behavior in the non-decoupling situation, we show in Fig.~\ref{fig:FTrunning_NonDecoupling} (right) the running of $\mu_\phi^2$ from the low scale up to high scale, and then back down after a variation at the high scale. 
With only a 1\%-level variation at the high-energy scale, we see that the low-energy value of $\mu_\phi^2$ shifts from $-10^4$ GeV$^2$ to $-10^7$ GeV$^2$. This indicates that in order to obtain the desired low-energy value for $\mu_\phi^2$, a large amount of fine-tuning is necessary, which is expected in the naive approach to the CW potential with large hierarchies. When using this procedure in the decoupling approach, one has to use the corresponding RGEs (\ref{run_dec_lambda}) and (\ref{run_dec_muphi}).  The effect of a variation at the high scale will be smaller compared to the non-decoupling approach, as the mass thresholds reduce the impact of the heavy degree of freedom on the low-energy parameter. 

\begin{figure}[t!]
    \centering
    \includegraphics[scale=0.76]{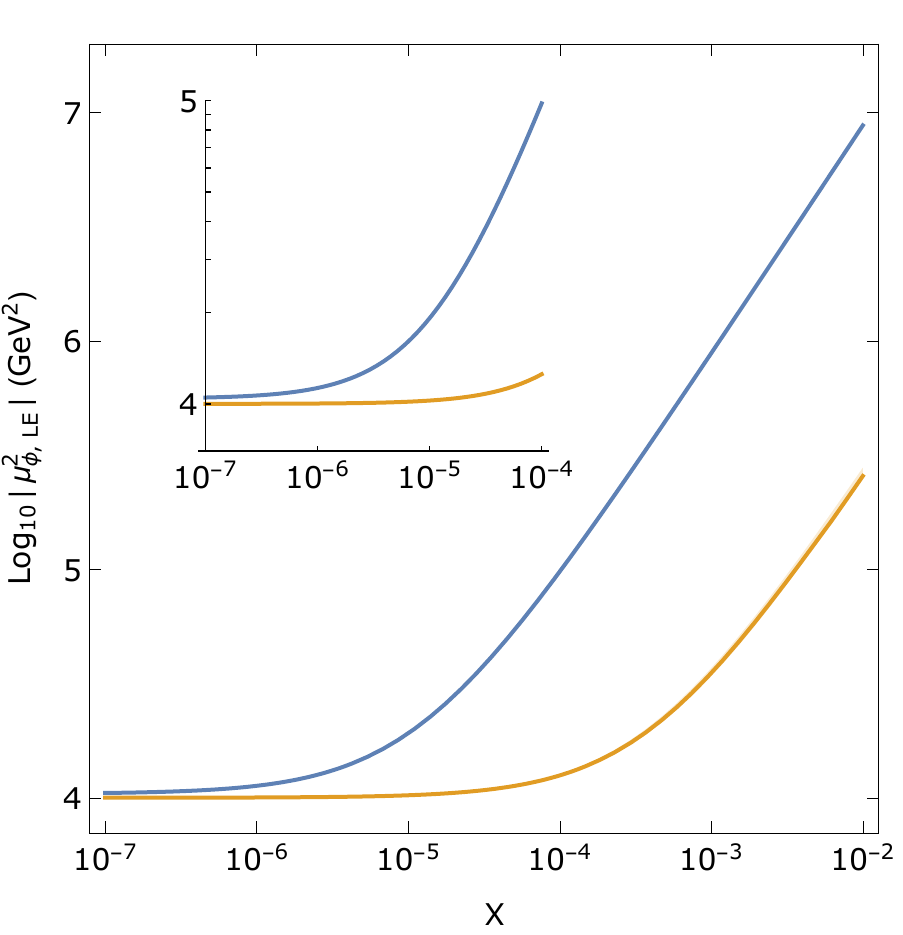}
    \caption{The effect of small variations in the high-energy value of the parameters on the low-energy value  $\mu_{\phi,\text{LE}}^2 = \mu_\phi^2(\mu_\text{min})$, using the benchmark point with $\alpha > 0$. The results for the decoupling (orange line) and non-decoupling (blue line) cases are compared. A field value of $\phi_c = 300$ GeV is used.}
    \label{fig:FTcomparison}
\end{figure}

\begin{figure}[t!]
    \centering
    \includegraphics[scale=0.75]{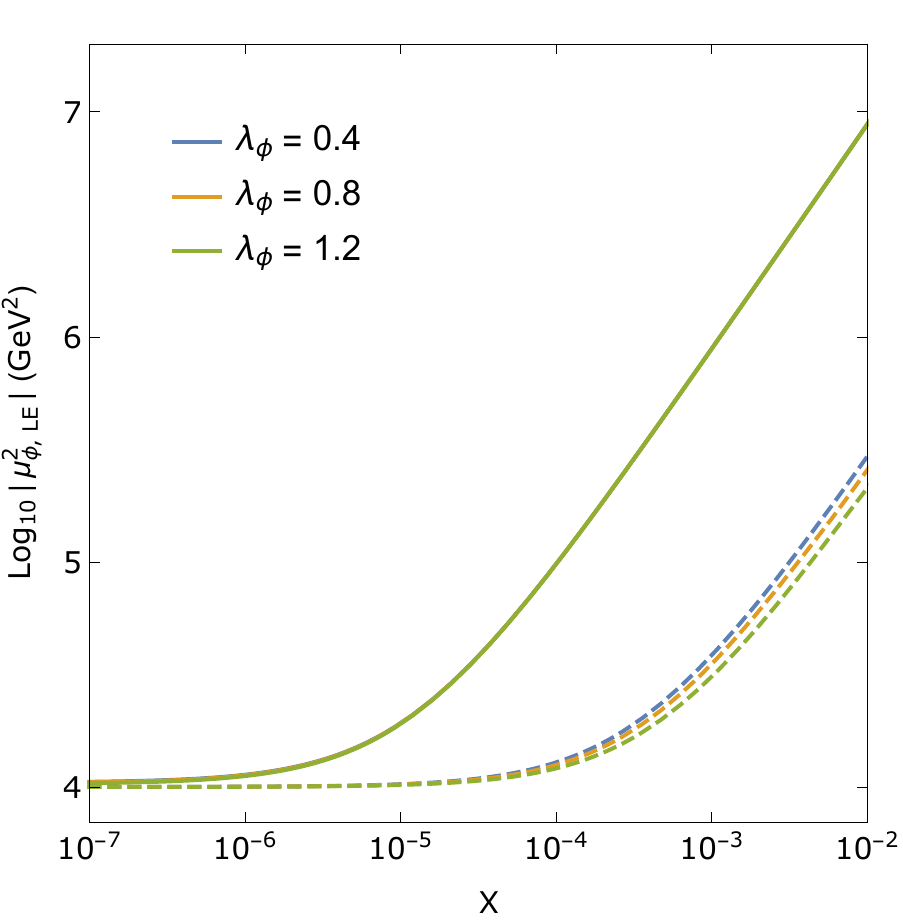}
    \hspace{0.5 cm}
    \includegraphics[scale=0.75]{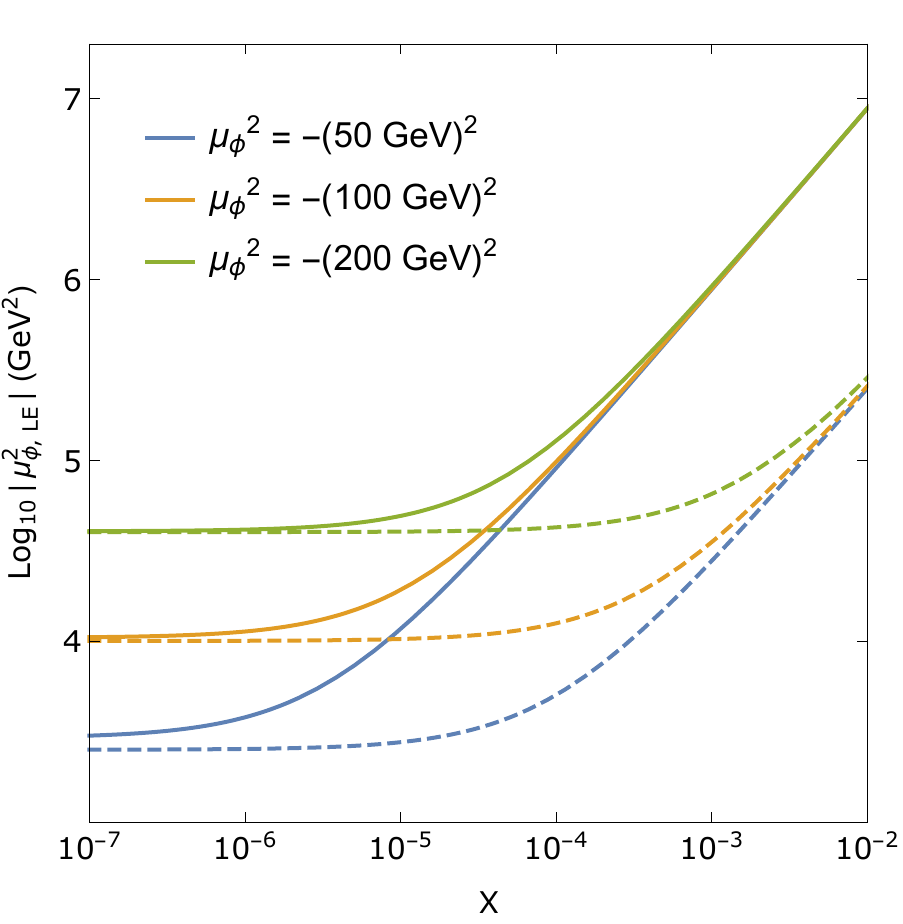}\\
    \vspace{0.5 cm}
    \includegraphics[scale=0.75]{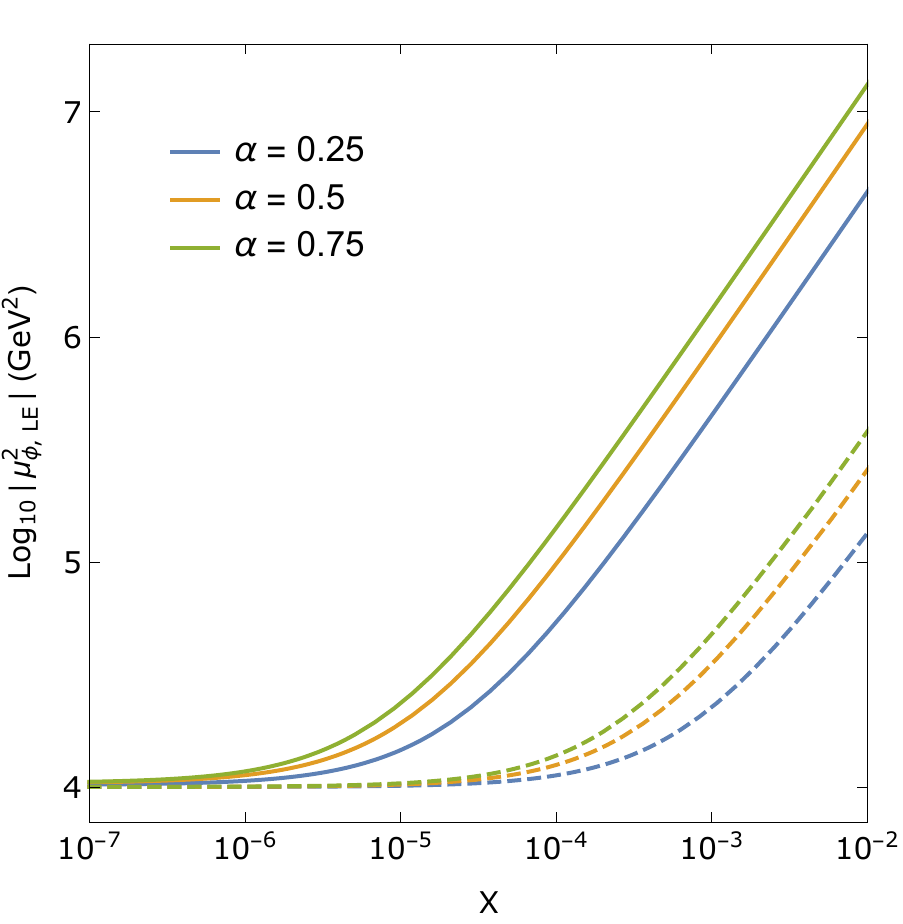}
    \hspace{0.5 cm}
    \includegraphics[scale=0.75]{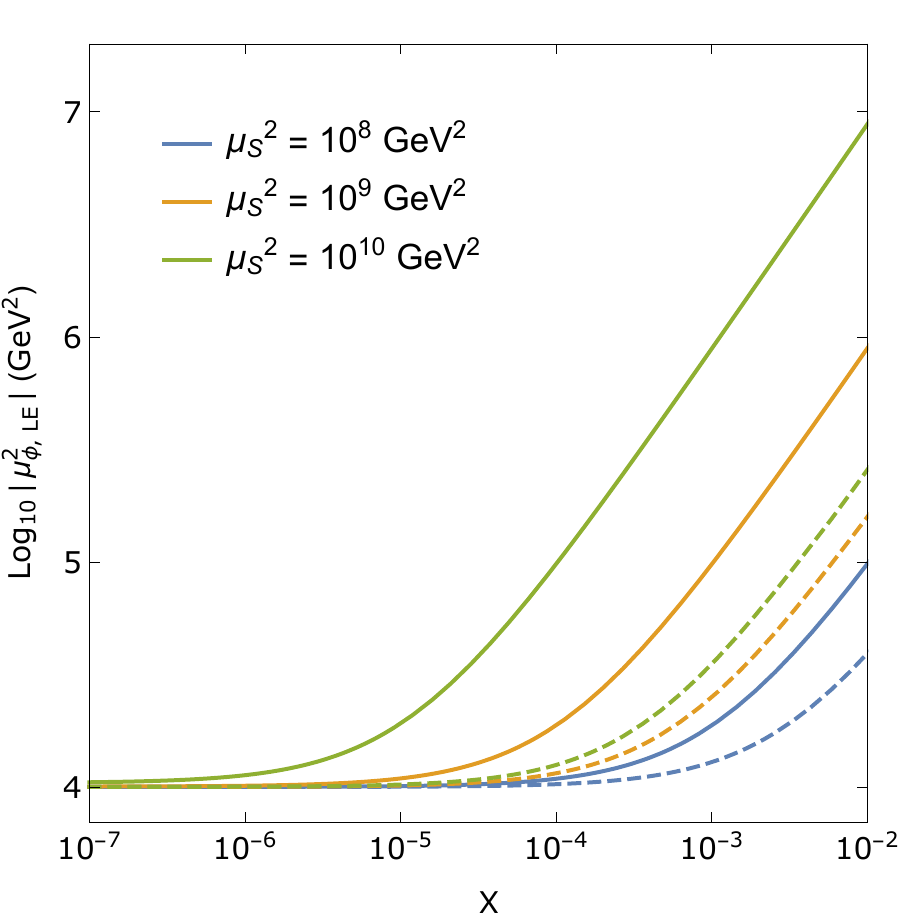}
    \caption{The low-energy value $\mu_{\phi,\text{LE}}^2 = \mu_\phi^2(\mu_\text{min})$ as a function of $X$ for different model parameter values. The fixed parameters have the value of the benchmark point in Eq.~ \eqref{banch} with $\alpha > 0$. The solid lines show the results in the non-decoupling case, the dashed lines show the decoupling results.}
    \label{fig:ft_vary_parameters}
\end{figure}
In order to compare the sensitivity to the high-energy parameters with and without the decoupling prescription, and to be more quantitative, we show in Fig.~\ref{fig:FTcomparison} the value of $\mu_\phi^2$ evaluated at $\mu = \mu_\text{min}$ as a function of $X$, respectively in the solid orange and solid blue line. One observes that $\mu_\phi^2(\mu_\text{min})$ is considerably less sensitive to variations at the scale $\mu_\text{max}$ when using the decoupling method.  However, there is still quite some fine-tuning necessary even when using the decoupling method, since for a percent-level variation of the high-energy parameters, $|\mu_\phi^2|$ changes more than one order of magnitude. To illustrate the robustness of these findings, we explore the behavior of $\mu_\phi^2$ when varying the model parameters $\lambda_\phi$, $\alpha$, $\mu_\phi^2$ and $\mu_S^2$. We aim both to study how the fine-tuning depends on these parameters, and to show that the choice for our benchmark point is not special. In Fig.~\ref{fig:ft_vary_parameters}, the value of $\mu_\phi^2$ at $\mu = \mu_\text{min}$ is given as a function of $X$. In each plot, a different parameter is varied one at a time, while the remaining ones are kept at the same value as in the benchmark point, with $\alpha > 0$ in all plots (negative $\alpha$ yielding qualitatively similar results). Both the decoupling (dashed lines) and non-decoupling (solid lines) results are shown.  In the upper-left panel, there is almost no dependence on the value of $\lambda_\phi$ in the decoupling case  (solid lines). Therefore, the three lines overlap almost completely, and only the result for $\lambda_\phi = 1.2$ is visible.

We end this section by noting that the large sensitivity at low energy to changes of the parameters of the model at high energy is very different from the analogous sensitivity to the standard logarithmic RG running one has in any theory. Running the SM parameters up to the GUT scale for instance and then changing the parameters at the GUT scale by a little and then running down, will also generate a substantial difference at low energies. The difference is that in the present case the effects depend quadratically on the large scale as opposed to logarithmically.
\section{Alternative implementations of the decoupling method}
\label{sec:comparison}
The decoupling theorem suggests that heavy degrees of freedom cannot be excited, and hence cannot affect the physics, at energies smaller than their masses. However, it is not an exact recipe that specifies at which particular scale the decoupling happens. In the former section, we chose perhaps the simplest prescription \cite{Casas:1998cf}, where the decoupling scale is fixed completely by the (scalar) particle masses, the exact definition is $\mu_i = m_ie^{-3/4}$. 

\begin{figure}[t!]
    \centering
    \includegraphics[scale=0.8]{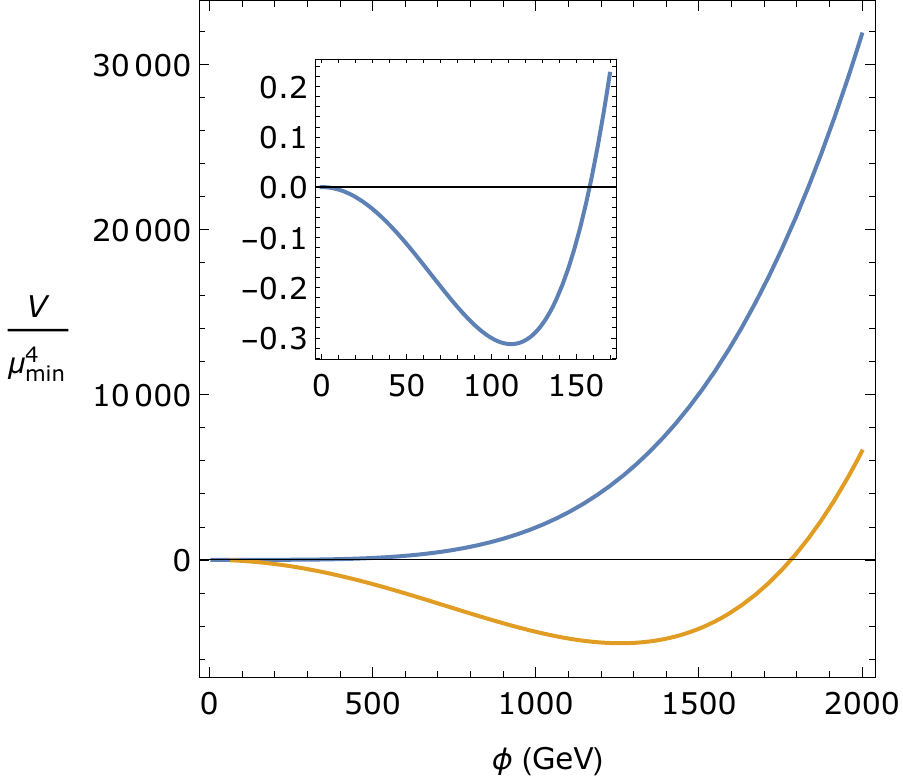}
    \caption{The shape of the potential in the tree-level case at $\mu_{\min}$ (blue) and using the decoupling method Eq.~\eqref{eq:Veff_ren_dec_multi} (orange) for $\alpha=0.5$. The decoupling scales are set to $\mu_{m_i} = 1.01 \cdot m_i e^{-3/4}$.} 
    \label{fig:GeneralDecoupling}
\end{figure}

In ref.\ \cite{Casas:2000mn}, a generalization of the decoupling method for the effective potential is put forward. The main modification is the appearance of decoupling scales, $\mu_{m_i}$, that are in general different from the mass thresholds $m_i$.  In this approach the one-loop potential in our model would read
\begin{equation}
    V^{(1)}_{\textrm{dec}} = \frac{1}{4(4 \pi)^2} \sum_{i=\phi,S} m_i^4 \left[ \log \frac{m_i^2}{\mu_{m_i}^2} +\theta(\mu-\mu_{m_i}) \log \frac{\mu_{m_i}^2}{\mu^2} -\frac{3}{2}\right] \, .
    \label{eq:Veff_ren_dec_multi}
\end{equation}
For $\mu_{m_i}= m_i e^{-3/4}$ the expression in Eq.~\eqref{eq:Veff_ren_dec} is recovered. When $\mu$ is larger than all the $\mu_{m_i}$, Eq.~\eqref{eq:Veff_ren_dec_multi} corresponds to the effective potential Eq.~\eqref{eq:Veff_ren} with no decoupled particles. As the authors in ref.\ \cite{Casas:2000mn} suggest, a reasonable choice for the decoupling scales is $\mu_{m_i} \simeq m_i$, as this improves the validity of the perturbative expansion and follows the reasoning of the decoupling theorem \cite{Symanzik:1973vg,Appelquist:1974tg}. As mentioned, there is no exact recipe that specifies at which particular scale the decoupling happens, but we point out that 
choosing slightly different scales for $\mu_{m_i}$ can have a large effect on the shape of the potential.
In order to make our point, we consider two different choices for the decoupling scales to be inserted in Eq.~\eqref{eq:Veff_ren_dec_multi}, namely $\mu_{m_i}= m_i e^{-3/4}$, which is the value of the decoupling scale in the potential \eqref{eq:Veff_ren_dec}, and $\mu_{m_i}=1.01 \, m_i e^{-3/4}$. 
In the first case, as shown in Fig.~\ref{fig:RGrunningCombined} (right), we obtained a very good agreement between the tree-level and one-loop potential. In the latter case, despite the decoupling scales being changed at the percent level, the one-loop correction shifts the minimum far away from the tree-level value (more than one order of magnitude), as clearly shown in Fig.~\ref{fig:GeneralDecoupling}.\footnote{The change is not so dramatic if one compares pairs of scales different from $\mu_S e^{-3/4}$, say $\mu_{m_i}=1.01 \, m_i$ and $\mu_{m_i}=1.02 \, m_i$. However, in that case the VEVs are both much larger than the tree-level result.}  The origin of the large sensitivity is the residual log-term  (the first term in the squared brackets in Eq.~\eqref{eq:Veff_ren_dec_multi}), which is multiplied by the large scale $m_S^4 \sim \mu_S^4$. This term gives a large contribution to the potential for $\mu_{m_S} \neq m_S e^{-3/4}$. It is worth mentioning that if one takes decoupling scales $\mu_{m_i} < m_i e^{-3/4}$ (even slightly smaller), the minimum of the one-loop corrected potential disappears because the overall sign of the residual large logarithm in Eq.~\eqref{eq:Veff_ren_dec_multi} is negative. This applies when $\alpha>0$, while the situation is reversed when taking $\alpha<0$, so in that case symmetry breaking will only occur when $\mu_{m_i} < m_i e^{-3/4}$. Hence, just like the non-decoupling approach discussed in Sec.~\ref{sec:properties}, the effective potential in Eq.~\eqref{eq:Veff_ren_dec_multi} leads to a strong dependence of the low-energy physics on the sign of $\alpha$, i.e.~on the interactions involving the high-energy scalar. This behavior was not present when using the decoupling implementation discussed in Sec.~\ref{sec:decoupling} according to Eq.~\eqref{eq:Veff_ren_dec}.
Our conclusions on the large sensitivity at low energy to changes of the parameters of the model at high energy still hold when considering the more general decoupling approach \eqref{eq:Veff_ren_dec_multi}, since the RG running of the parameters is only slightly altered compared to the RG running described in Sec.~\ref{sec:decoupling}. 

It is clear from the above discussion that the extreme sensitivity to changes in the parameters of the theory appears only when one considers changes of parameters above the decoupling scale. Usually it is natural to define parameters at or around the scale at which they appear. For example, SM parameters are often defined at the $M_Z$, whereas GUT parameters would be fixed at the GUT scale. Since the decoupling scale is of the order of the scale of the high-energy theory, one could  avoid the fine-tuning problem by supplementing the decoupling method with the requirement that all parameters of the theory are fixed at or below the decoupling scale of the high-energy scalar. In that way, there will be no large sensitivity to changes of the parameters at this scale. Running downward will not lead to large changes, while running upward will also not be very dependent on variations of parameters at  the scale where the parameters are defined (we have numerically checked this statement). The fine-tuning problem that we have addressed arises from running upward beyond the decoupling scale, then modifying the parameters at some higher scale and running down again. In the case of one decoupling scale at high energies, this prescription that all parameters of the high-energy theory should be fixed at or below the decoupling scale offers a way to avoid the fine-tuning problem. When there are multiple decoupling scales, e.g.\ in case the sum in Eq.~(\ref{eq:Veff_ren_dec_multi})  includes a third particle with a mass that is substantially higher than $m_S$, there will be large sensitivity to changes of the parameters of the theory at the highest of the decoupling scales (in practice a factor 2 higher will already give rise to a fine-tuning problem). Therefore, in order to avoid the fine-tuning problem one should fix the parameters at or below the decoupling scale of the degrees of freedom associated to them. Needless to say, this is just a way to avoid in practice the extreme sensitivity to changes in the parameters.

\section{Conclusion}
\label{sec:concl}
In this paper we have carried out an analysis of the hierarchy problem from the viewpoint of the effective potential. Our study does not propose any solution to the hierarchy problem, rather it aims to show that this issue is present in the extraction of the observables from the effective potential, even when using a decoupling approach. For this purpose we have mainly concentrated on the decoupling method of \cite{Casas:1998cf}, which freezes the effect of heavy particles on the running at low energies. Working at the one-loop level, and within a simple two-scalar theory, we disentangled the effects of the high-energy degree of freedom on the shape of the potential and on the fine-tuning of the model parameters. We find that while the decoupling method leads to an acceptable and convergent effective potential, the method does not solve the fine-tuning problem that is inherent to the hierarchy problem of multiple-scale theories. Compared to the non-decoupling approach, the amount of fine-tuning is considerably reduced. Nevertheless, the mass of the low-energy scalar turns out to be still very sensitive to small changes of the parameters at the scale of the high-energy sector.

We also considered another implementation of the decoupling method which leads to different conclusions on the impact of quantum corrections on the shape of the potential, in particular on the VEV of the low-energy scalar field. For the decoupling potential in Eq.~\eqref{eq:Veff_ren_dec_multi}, we find that small deviations from the ``reference'' decoupling scale $\mu_S e^{-3/4}$ make the minimum shift far away from the tree-level value if $\mu_{m_i}>m_i e^{-3/4}$, whereas the minimum is absent whenever $\mu_{m_i}<m_i e^{-3/4}$, when $\alpha>0$. The situation is reversed when negative values for $\alpha$ are used. Therefore, the decoupling method outlined in Eq.~\eqref{eq:Veff_ren_dec_multi} reintroduces the dependence of the shape of the potential, and the existence of a non-trivial minimum, on the sign of the coupling $\alpha$ (as it was for the non-decoupling approach described in Sec.~\ref{sec:potential_no_dec}).  The fine-tuning problem is very similar in the two implementations, however. 

We ended by suggesting a way to avoid running into this fine-tuning problem in such decoupling approaches by adopting the prescription to fix parameters at or below the decoupling scale of relevance to them.    We emphasize that this is a prescription to avoid running into the fine-tuning problem, not to remove it altogether.

\section*{Acknowledgements}
The work of S.B. is supported by the Swiss National Science Foundation (SNF) under the Ambizione grant PZ00P2\_185783. S.B. is grateful to the FSE fellowship program at the University of Groningen where this work was started. The work of R.P. is supported by the NWO programme ``Higgs as a probe and portal". The authors thank Bogumi\l a \' Swie\. zewska for useful discussions.
\newpage
\bibliographystyle{hieeetr}
\bibliography{paper.bib}
\end{document}